\documentclass[twocolumn,3p,times]{elsarticle}
  
\usepackage{mathptmx, courier, pifont}
\usepackage[scaled=0.92]{helvet}
\usepackage[T1]{fontenc}
\usepackage{textcomp}
\usepackage{amsfonts}

\usepackage{epsfig}
\usepackage{amssymb}
\usepackage{graphicx}
\usepackage{amssymb}
\usepackage{graphicx}
\usepackage{mathtools, cuted}
\biboptions{numbers,sort&compress}
\usepackage[table]{xcolor}

\begin{document}

\begin{frontmatter}

\title{Monte Carlo simulation of COVID-19 pandemic using statistical physics-inspired probabilities}

\author[a]{Jos\'e Enrique Amaro}
\ead{amaro@ugr.es}
\ead[url]{webpage: http://www.ugr.es/$\sim$amaro}

\author[b]{Jos\'e Nicol\'as Orce*}
\ead{jnorce@uwc.ac.za}
\ead[url]{https://nuclear.uwc.ac.za}

\address[a]{Departamento de F\'isica At\'omica, Molecular y Nuclear and Instituto Carlos I de F\'isica Te\'orica y Computacional, 
              Universidad de Granada, E-18071 Granada, Spain}

\address[b]{Department of Physics \& Astronomy, University of the Western Cape, P/B X17 Bellville ZA-7535, South Africa}

\date{\today}

\begin{abstract}

  We present a  Monte Carlo simulation model of an epidemic spread inspired on physics variables such as temperature, 
  cross section and interaction range, which considers the Plank distribution of photons in the black body radiation to describe 
  the mobility of individuals. The model consists of a
  lattice of cells that can be in four different states: susceptible,
  infected, recovered or death.  An infected cell can transmit the
  disease to any other susceptible cell within some random range $R$.
  The transmission mechanism follows the physics laws for the
  interaction between a particle and a target. Each infected particle
  affects the interaction region a number $n$ of times, according to
  its energy.  The number of interactions is proportional to the
  interaction cross section $\sigma$ and to the target surface density
  $\rho$. The discrete energy follows a Planck distribution law, which
  depends on the temperature $T$ of the system.  For any interaction, 
  infection, recovery and death probabilities are applied. We investigate the results of viral transmission 
  for different sets of parameters and compare them with available {\small COVID-19} data. The
  parameters of the model can be made time dependent in order to
  consider, for instance, the effects of lockdown in the middle of the
  pandemic.
  
\end{abstract}

\end{frontmatter}

\section{Introduction}

The {\small COVID-19} pandemic, that has so far caused almost five million deaths worldwide~\cite{corona1,corona2}, has recently induced 
numerous studies on the mathematical modelling of temporal distributions of infected 
cases and fatalities \cite{Gar21,Coo20a,Coo20b,Lan21,Pur21}. Various statistical models for pandemic spread and their predictive power 
have been benchmarked against data~\cite{anydist,amaro1}.

The {\small SIR} (susceptible-infected-recovered) model -- developed by Kermack and McKendric~\cite{KM} -- and derived models are the most
widely used to study viral spreading of contagious epidemics or mass immunization planning~\cite{Wei13,vaccine,vaccine2} since they provide mean
values of the cumulative incidence as a function of time in a deterministic way. {\small SIR} models belong to the compartmental type, where
the equations only involve the time variable, and the incidence functions refer to the total number of individuals in each compartment
or subset of the total number of individuals in the system. 

Generalization to more realistic models of the infection transmission process requires solving spatio-temporal equations~\cite{st} in a
two-dimensional space representing the surface of a region under
epidemics. Such individual-level models were first studied by Kendall
\cite{kendall1957} and Bartlett \cite{bartlett1957,bartlett1957_2},
who also considered both deterministic and stochastic kind of
descriptions.

In this paper we present a study of space-time propagation of a viral infection
using a new stochastic model inspired by concepts of statistical
physics and nuclear physics.  This model is compared to a simple compartment 
model developed in ref. \cite{amaro1} (the D model), where we studied the accumulated fatalities for a
series of countries during the first {\small COVID-19} wave.

The purpose of this work is to present the Monte Carlo Planck ({\small MCP})
model in order to study the stochastic behavior of the spatio-temporal
propagation, to test the predictive power of these kind of models with respect to {\small COVID-19} available data as well as 
to assess the validity of the D model.

Further Monte Carlo-type studies have been developed to simulate the pandemic transmission~\cite{nature1,mc2,Tri21}, but use other approaches 
such as, for instance, a 2D-random walk Monte Carlo simulation based on proximity infection spread. In the present {\small MCP} model we take a 
physical point of view,  where the infection is a result of an interaction, and it follows to some extent statistical laws similar to the 
interaction in a system of many particles. Thus we model the infection as a likely result of a physical interaction.
We apply concepts taken from  statistical and nuclear physics such as interaction range, energy,
temperature, and interaction cross section.
These concepts,  adapted to the present case of interest,  are useful to describe
probability distributions and interaction coefficients in physical
systems, which are here characterized by epidemiological probabilities
of infection, recovery and death.

In Section 2 we present the {\small MCP} model. In Section 3 we present results of the simulation applied to {\small COVID-19} 
fatalities data for several countries. Finally in Sect. 4 we draw our conclusions.

\section{The Monte Carlo Planck model}

  We consider the system of a two-dimensional lattice or grid, where
  each unit cell has two coordinates $k = (k_1,k_2)$, with
  $k_i=1,2,\ldots,L$, where $L$ is the total size of the system, with
  surface $L^2$. Each cell $k$ is occupied by an individual, who can
  be in any of four states or classes: susceptible, infected,
  recovered or dead. This state is specified by four fields or
  matrices $s(k)$, $i(k)$, $r(k)$, and $d(k)$ that can take the values
  1 or 0 (state quantum numbers) , depending whether the individual in
  cell $k$ belongs or not to the corresponding class. The model can be
  easily extended to include other classes or intermediate states.

  In the initial state, for  $t=0$, all the individuals are susceptible. Then
  \begin{equation}
    s(k)=1 ,  \kern 1cm i(k)=r(k)=d(k)=0
  \end{equation}
  for all $k$ except for one initially infected cell $k_0$, where
  \begin{equation}
    s(k_0)=0, \kern 1cm i(k_0)=1.
  \end{equation}
  We choose the initial infected cell at the center of the grid, but
  in the model it can be chosen randomly or even to be several infected cells.

  In the simulation we compute the state of the system in time
  intervals of $\Delta t=1$ day.  At the end of the day, $t=1,2,3,\ldots$, we
  compute and store in the matrices $s$, $i$, $r$, $d$ the current
  state of the system. We also store the total number of susceptible
  $S(t)$, infected, $I(t)$, recovered, $R(t)$ and dead, $D(t)$ and the
  daily increments, $\Delta S(t)$, $\Delta I(t)$, $\Delta R(t)$,
  $\Delta D(t)$, and repeat the calculation for the next day.

  In the simulation for day $t$ we apply the following algorithm
  to each cell $k$.

\begin{enumerate}

\item   If it is not infected, $i(k)=0$, do nothing.

\item It it is infected, $i(k)=1$, then we assume that it can infect
  only people in other cells within a finite region of size $R$.  
  The value of the range $R$ takes into account the zone of
  influence of individuals in one day.  To simplify the algorithm in
  Cartesian coordinates we assume that the region is a square of side
  $2R$, centered at the cell $k$, but it could be also circular or
  other shape.

  Now the Monte Carlo starts by computing the value of $R$ randomly 
  with some probability distribution. We assume an exponential law
  \begin{equation}
    P(R)= \frac{{\rm e}^{-R/R_0}}{R_0}, 
  \end{equation}
  which indicates that is less probable to move far away from the average
  interaction range $R_0$ of the individuals. This is one of the
  parameters of the model.

\item Once $R$ is chosen for an infected particle, it can interact
  randomly with any of the $N_R$ individuals within the region with
  area $S_R$.  As in particle physics, we characterize the interaction
  between individuals by a cross section $\sigma$.  The number of
  possible interactions is then given by the probability formula when
  $n$ particles are shot over a target
  \begin{equation}
    N_{\rm int} =  n  N_R \frac{\sigma}{S_R} =
    n    \rho \sigma,
    \end{equation}
  where $\rho=1$ is the surface density and the number of shots $n$ is
  related with the mobility of the individual within its zone of
  influence, indicating if it has a narrow social behaviour and how
  often it moves around each day. We call this property the {\em
    energy} of the individuals. For instance, Young people have more
  energy than old people, or people with a job where they interact
  often with other people. We compute the value of the energy $n$
  randomly with an energy distribution. In analogy to the Planck law
  for photons from statistical physics, we write the energy
  distribution in a simplified form as
  \begin{equation}
    P(n)= \frac{{\rm e}^{-n/T}}{{\rm e}^{1/T}-1}, 
    \end{equation}
  where the parameter $T$ is the {\em temperature} of the system
  that determines the average energy or number of shots.

  In this work we choose unit cells and therefore assume $\rho=1$ in
  arbitrary units, but the model can be modified to include specific
  densities and sample population sizes with known number of
  individuals. In the model the information on the density is included
  in the range $R_0$. To increase the density is equivalent to
  increase the range and the temperature. The result of this work are
  given for unit density and the values of $R_0$ and $T$ should be
  appropriately re-scaled to apply the results to different densities.

\item Once we know the range $R$ and the number of interactions
  $N_{\rm int}$ of an infected cell, we choose randomly $N_{\rm int}$
  cells, with coordinates $k'$,
  within the interaction region. If cell $k'$ is susceptible,
  $s(k')=1$, it can be infected with probability $0 \leq p_i \leq
  1$. We then decide randomly it it is infected or not according to
  that probability. In case that it becomes infected, we change the
  vaue of the corresponding matrix elements $s(k')=0$ and
  $i(k')=1$. We also store the time it becomes infected in the
  infection time matrix $\Theta(k)=t$.  This matrix is important for
  the nex step in deciding when the particle becomes recovered or dead.

  \item After interacting with all the $N_{\rm int}$ individuals
    within the range $R$, along the day, finally we decide if cell
    $k$ becomes recovered or dead  at night. We assume that the removal
    probability depends on time with a function
    \begin{equation}
      P_{\rm rem}(t)= \frac{1}{1+{\rm e}^{(t_r-t)/b}},
    \end{equation}
    where $t_r$ is the time such that $P(t_r)=1/2$  and 
     $t_r \gg b$, so that the removal probability is very
    small for low times, and later it increases with time at a rate
    driven by the parameter $b$. For $t\rightarrow\infty$, the removal
    probability is one. All the individuals recover or die.

    Therefore we randomly decide if $k$ is removed at the end of the
    day with a probability $P_{\rm rem}(t-\Theta(k))$ that depend on
    the time $t-\Theta(k)$ that particle $k$ has been infected.
    In case $k$ is removed, we change the state quantum number in
    the infected matrix, $i(k)=0$.

  \item The final step, if particle is removed, we decide if it is
    dead randomly with probability $P_{\rm d}\leq 1$. Otherwise it is
    recovered. Finally we store the matrices $r(k)=1$ (0), and
    $d(k)=0$ (1) accordingly.
    
\item We repeat the procedure on the next cell.

\end{enumerate}

\begin{figure*}[t]
\begin{center}
\includegraphics[width=6.5cm,height=5cm,angle=-0]{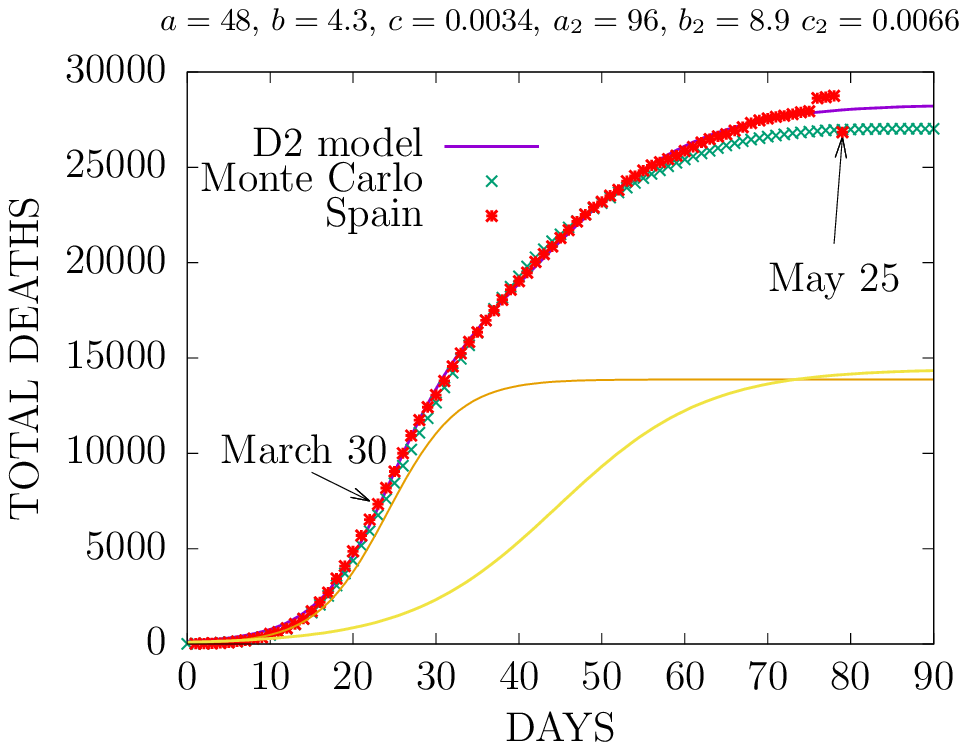}
\hspace{0.2cm}
\includegraphics[width=6.5cm,height=5cm,angle=-0]{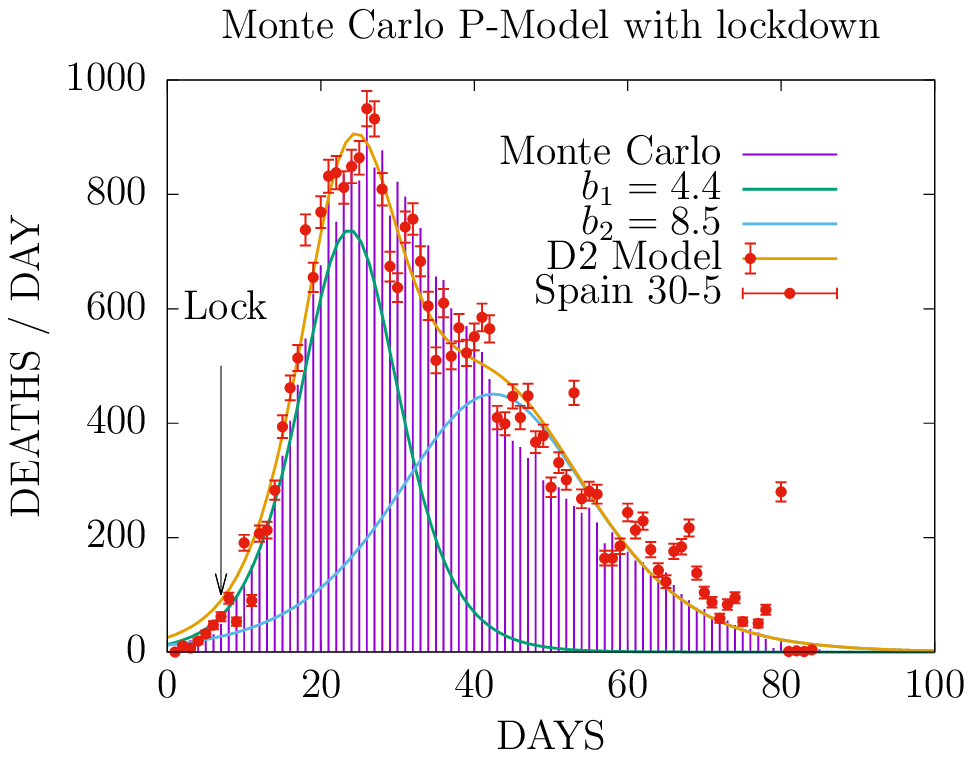}
\caption{Fit of the P-model to the cumulative deaths and deaths per day by {\small COVID-19} in  Spain  during the first wave. The red points are the data compiled up to May 25, 2020. The two yellow lines in the left panel are the two D-functions 
appearing in the D2 model, Eq. (\ref{Dmodel}).
}
\end{center}
\end{figure*}

The algorithm so defined depends on the following parameters
    
  \begin{itemize}
  \item The size of the system $L$, The unit of length in natural
    units, is the average distance between individuals. This
    determines the number of cells $L^2$.

  \item The range of interaction $R_0$, indicating the maximum
    distance on average that the infection can propagate in one day. This
    parameter can be time dependent, and, by changing it, one can simulate for
    instance lockdown or long travels in holidays.

  \item The cross section $\sigma$ measures the probability of
    interaction between two individuals. In the case of classical
    particles interacting with contact forces the cross sections is
    just the effective geometrical area of the pair.  For long range
    forces the cross section is larger.  In our case it is typically
    much larger than the geometrical human size, because this
    parameter takes into account the human behavior. Humans move and
    are social increasing its cross section, like a long range force.

  \item The temperature of the system, $T$ determines the average energy
    of the individuals, which is
    related to the movility within its range of interaction.

  \item  The infection probability $0\leq p_i \leq 1$

  \item The removal probability parameters $t_r$ and $b$ (two parameters).

  \item The death probability of removed $p_d$  
    
  \end{itemize}

\begin{figure*}[t]
\begin{center}
\includegraphics[width=6.5cm,height=5cm,angle=-0]{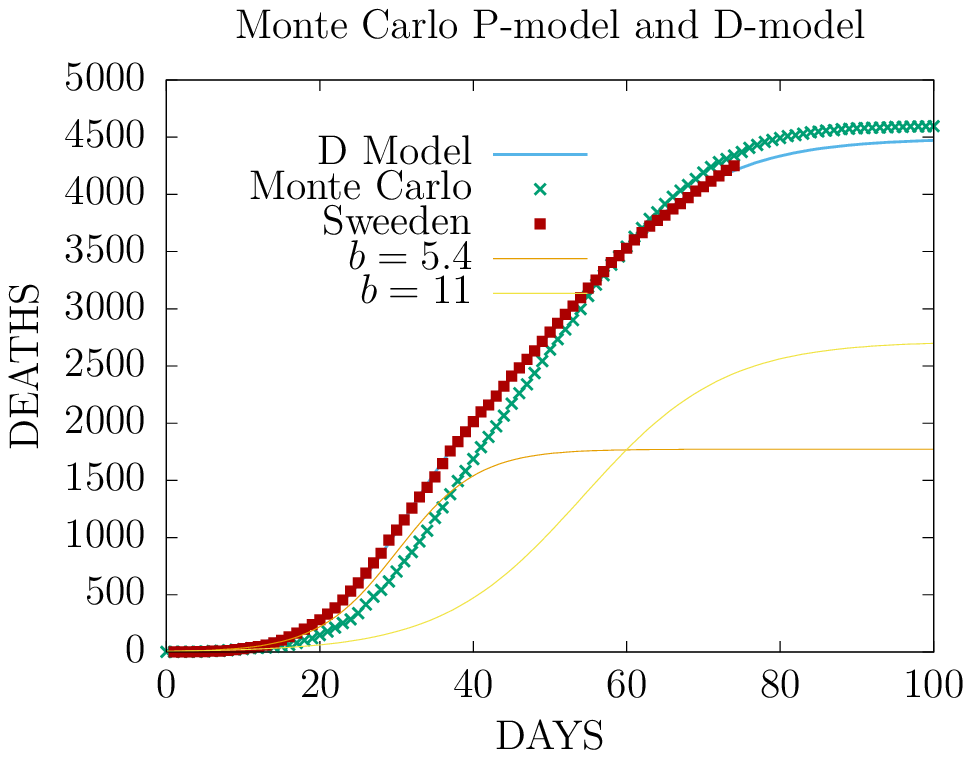}
\hspace{0.2cm}
\includegraphics[width=6.5cm,height=5cm,angle=-0]{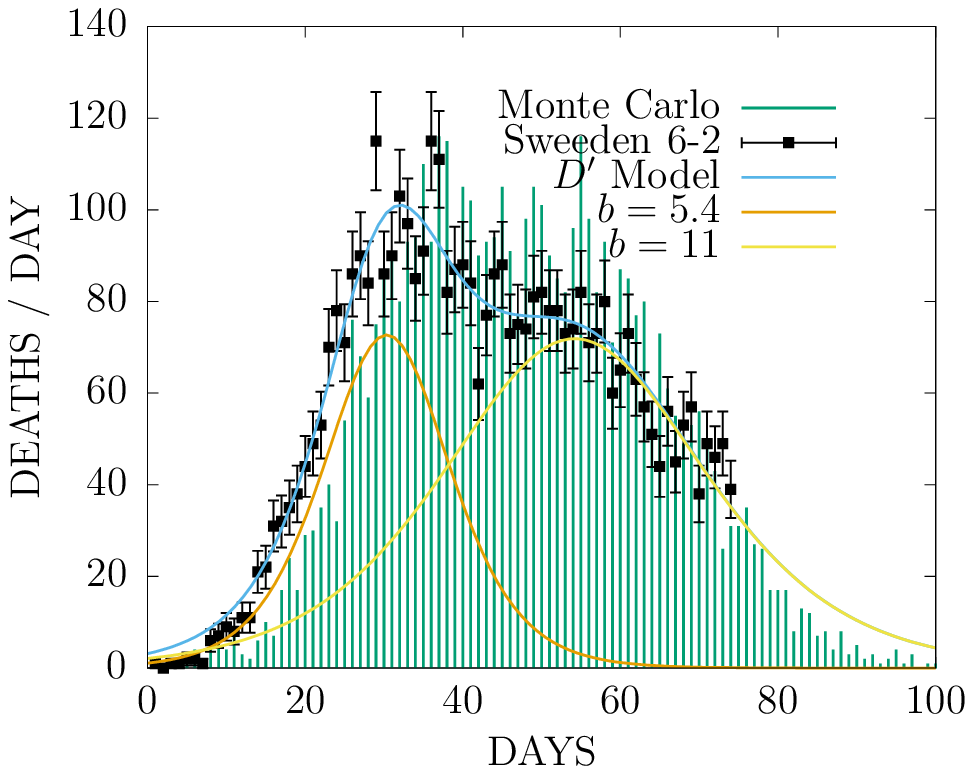}
\caption{Fit of the P-model to the cumulative deaths and deaths per day by {\small COVID-19} in Sweden  during the first wave.
}
\end{center}
\end{figure*}

These are in total 8 parameters $L$, $R_0$, $\sigma$, $T$, $p_i$, $a$,
$b$ and $p_d$. In addition, each one of these parameters can be easily
made time dependent to study the effects of political enforcement,
lockdown, social distancing, large events, etc.  The model could also
be modified easily to include coordinate dependence of the parameters
to specify for instance a large event in some region by increasing the
temperature or the cross section in that region.

Actually one can argue that increasing temperature, increasing cross
section and increasing the infection probability are expected to
produce similar effects on the pandemic evolution, and that the
pandemic evolution is effectively driven by less parameters, perhaps
six or less, unless time or space dependence (or non constant
densities) is introduced in them.

In the next section we fit the parameters of this {\small MCP} model to describe
 data from the first wave of the {\small COVID-19} pandemic.

\section{Results}

Firstly, the {\small MCP} model requires fixing the number of individuals, which is
equal to the area $N=L^2$ of the square lattice (each
individual occupies one square). To make the calculation manageable,
we chose an initial size of $L=300$, that is $N=90,000$ individuals in the sample,
although we latter also consider the effect of making L larger. Obviously,
if the number of deaths exceeds this figure, to study a more
realistic case, a size comparable to the population of the country in
question should be considered. But in this first simulation we prefer
to make the model as simple and manageable as possible in order to 
investigate whether under these simplifications it is still possible to describe the
data.

An important detail is that, given the probability of death $p_D$, the
expected mean number of deaths is fixed from the beginning 
$ D (\infty) \simeq N p_D $. 
If this value is not known experimentally, we
first fix it by fitting a {\small SIR}-type model to the incomplete
data. Specifically, we apply the D- and D2-models of
Ref. \cite{amaro1}, which consist of sums of logistic functions, to
estimate the probability of death before doing the Monte Carlo. The
death number as a function of time in the D-model is
\begin{equation} \label{Dmodel}
D(t) = \frac{a}{{\rm e}^{-(t-t_0)/b}+c},
\end{equation}
where the three parameters $a,b,c$ are fitted to data and the initial 
time $t_0$ is arbitrary. In the case of the D2-model we fit the sum of
two D-functions with six parameters $a_1,b_1,c_1,a_2,b_2,c_2$ in
total. This provides a method for estimating a prediction for the
total number of deaths $D(\infty)=a/c$ ($a_1/c_1 + a_2+c_2$ in the
case of the D2-model).  This value $D(\infty)$ is used as input to set the parameters of the 
Monte Carlo simulation since we know that the D model works well as a
starting point for pandemic forecasting~\cite{amaro1}.  Thus the Monte
Carlo will be useful to study the stochastic distribution of daily
cases, since the {\small SIR} models only provide the mean values without
statistical fluctuations.

\begin{table*}[ht]
\begin{center}
 \caption{Parameters of the {\small MCP} model for Spain, Sweden and South
  Africa.  The value of the range $R_0$ during lockdown
  in Spain and the temperature $T$ during gathering restrictions in
  Sweeden are in parentheses.}
  \label{tab:para}
  \vspace{0.2cm}
\begingroup\setlength{\fboxsep}{0pt}
\colorbox{gray!70!yellow!10}{%
\begin{tabular}{ccccccccc}
\hline
     & $L$ & $R_0$ & $\rho\sigma$ & $T $ & $t_r$  & $b$ & $p_i $ & $p_d $ \\\hline
     &     &       &          &      &  (days)&  (days) &  &   \\\hline 

{\bf Spain} & & & & & & & & \\ \hline
First wave      & 300 & 8 (3) & 0.1  & 20  & 20.7 & 4.5 & 0.2 & 0.3 \\ \hline
{\bf Sweden} & & & & & & & & \\ \hline
First wave      & 300 & 10  & 0.3  & 20 (15) & 24  & 4  & 0.2  & 0.3  \\ \hline
{\bf South Africa} & & & & & & & & \\ \hline
First wave      & 300 & 2.9 & 0.09  & 20  & 20 & 4 & 0.15 & 0.2 \\ \hline
Second wave     & 300 & 3 & 0.1  & 20  & 20 & 4 & 0.15 & 0.5 \\ \hline
\hline
\end{tabular}
}\endgroup
\end{center}
 \end{table*}

\begin{figure*}[t]
\begin{center}
\includegraphics[width=5.2cm,height=5.2cm,angle=-0]{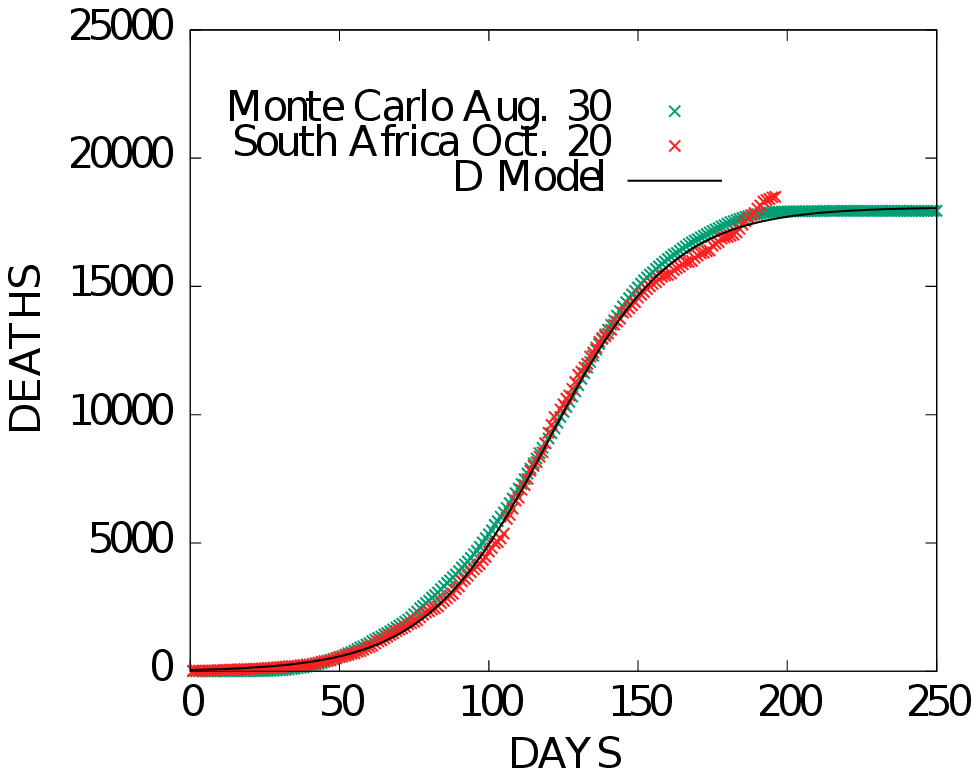}
\includegraphics[width=5.2cm,height=5.2cm,angle=-0]{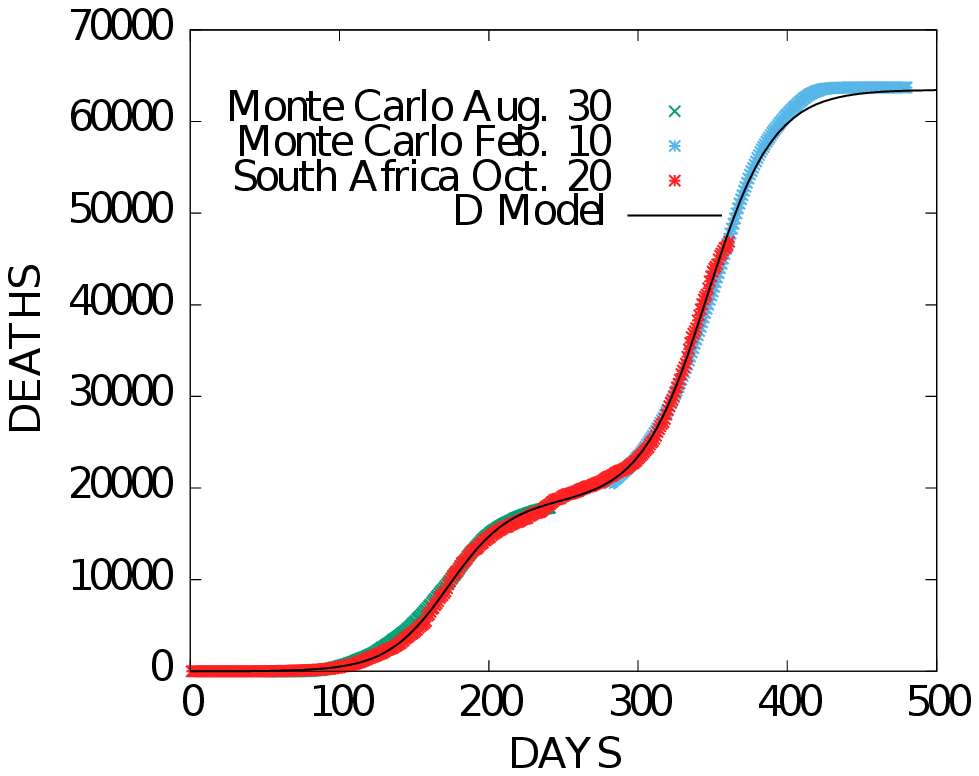}
\includegraphics[width=5.5cm,height=5.2cm,angle=-0]{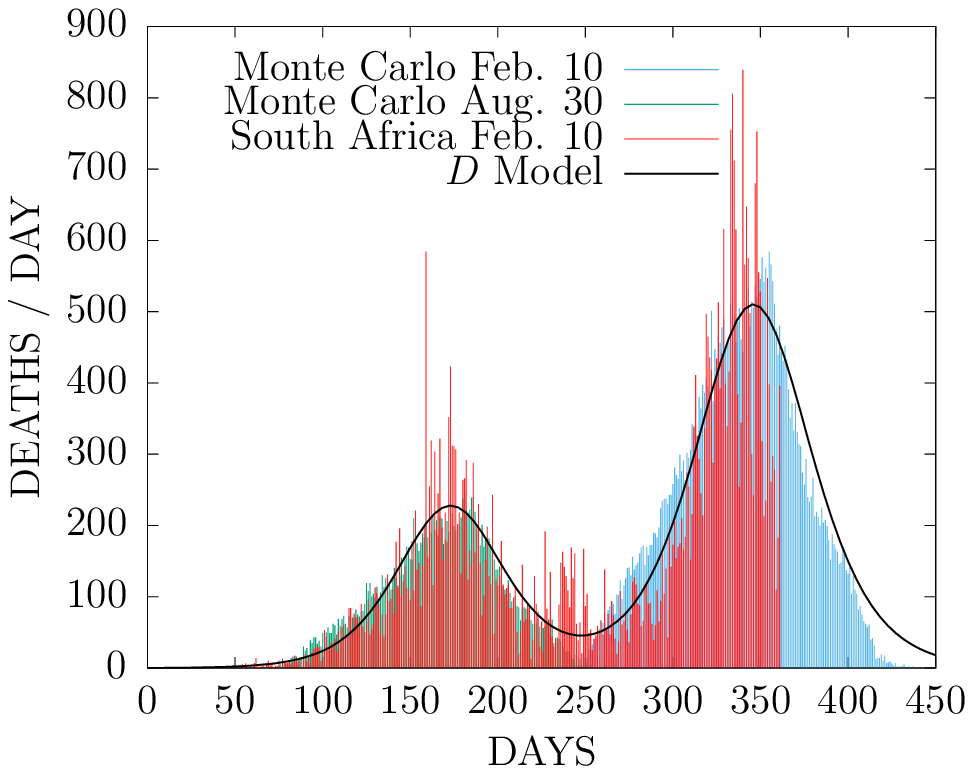}
\caption{ Fit of  {\small MCP} model to the cumulative deaths
  by coronavirus in South Africa during the first (left) and second
  (middle) pandemic waves, together with the overall daily deaths
  (right). Data from \cite{worldometer} fitted up to Feb. 10 2021.
}
\end{center}
\end{figure*}

The simulation was started assuming that in the initial instant $t=0$
there is an infected individual in the central cell of the array 
and all the others cells are susceptible. The algorithm of the previous
sections is then applied by assuming time steps $\Delta t= 1$ day.  In
Figure 1 we show the results of the Monte Carlo simulation of the
deaths-per-day, $\Delta(t)$, in Spain.  We fitted the data until May
30, 2020, where clearly the first {\small COVID-19} wave had practically ended.
We first fitted the D2 model (as the sum of two logistic functions) to
the data to better estimate the end of the epidemic, and then we
fitted the {\small MCP} paremeters to the curve of total accumulated deaths
$D(t)$ of the D2-model.  It is worth mentioning that fitting the Monte
Carlo to the accumulated deaths curve is more reliable than fitting
the daily data, $\Delta D(t)$, and gives better results.  This is
because the cumulative curve data has less fluctuations and less error
than the daily data.

The  {\small MCP} parameters are given in Table 1.  The parameters of Table 1
have been tuned to describe the experimental data, including the
lockdown. 
In general The Monte Carlo points have to be shifted in
time ($\Delta t=12$ days in the Spain case) to agree with the center
of the experimental peak, that has a different time origin. This is so
because there is an arbitrariness regarding the start of the pandemic.

 For Spain, the  {\small MCP} simulation includes lockdown effect on day 20, where the
 value of the range was reduced to $R_0=3$ (fitted also to the data).
 In Spain the lockdown started on day 14M (March 14), corresponding to
 day 7 in the plot of Fig.1, and the maximum of the daily deaths,
 $\Delta D(t)$, occurred on 2A (April 2), corresponding to day 26 in
 the plot. The distance between maximum and lockdown therefore is
 $t_{\max} -t_L = 26 - 7 = 19$ days.  In the the {\small MCP} simulation this
 distance is 15 days.\\

From the results of Fig. 1 we see that the {\small MCP} and the D2 models
present the same trend.  Both models describe the death data, both
total and daily deaths, very similarly.  This means that the
hypothesis of the {\small MCP} model that the spread of infection can be
described by physical interactions in a system of many particles is
correct, since it agrees with the statistical model of the {\small SIR} type
and is able to describe the experimental data of the pandemic.

The main difference with the D2 statistical model is that the {\small MCP}
acquires random statistical fluctuations, which the {\small SIR} models do not
have, since they only describe mean values. We also see that the
fluctuations in the data are much larger than in our model, even
considering the statistical errors of the daily data. These values 
have been taken from the Poisson distribution $\Delta n = \sqrt {n}$. 
However it must be taken into account that the official death data 
have systematic errors due to the data collection methods and
delays, and that these errors could be quite large.  In such cases it
may be reasonable to first compute a moving average of the data over three
or more days to reduce fluctuations before fitting the models.

\begin{figure*}
\begin{center}
\includegraphics[width=3cm,height=6cm,angle=-0]{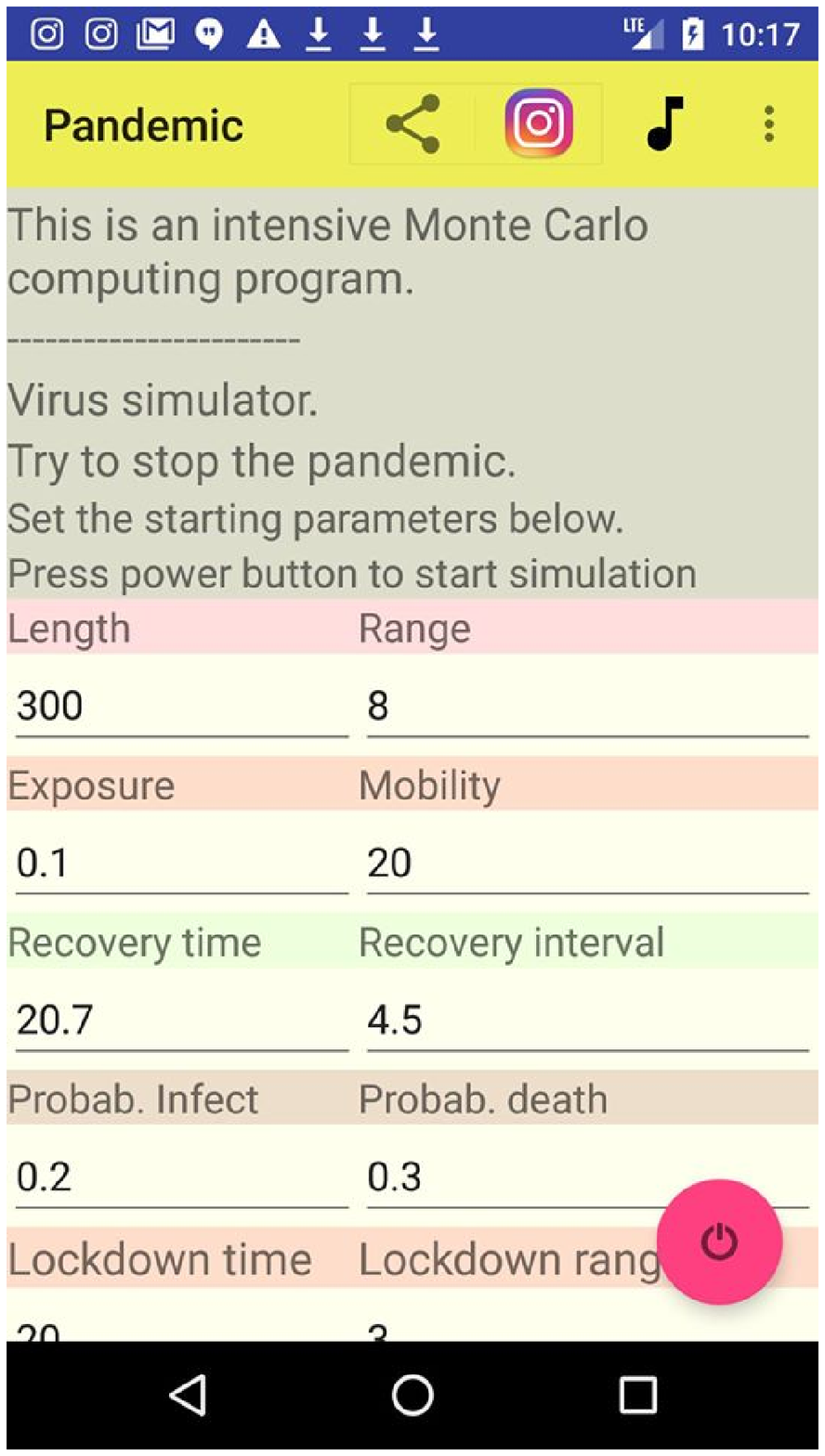}
\hspace{0.5cm}
\includegraphics[width=11cm,height=6cm,angle=-0]{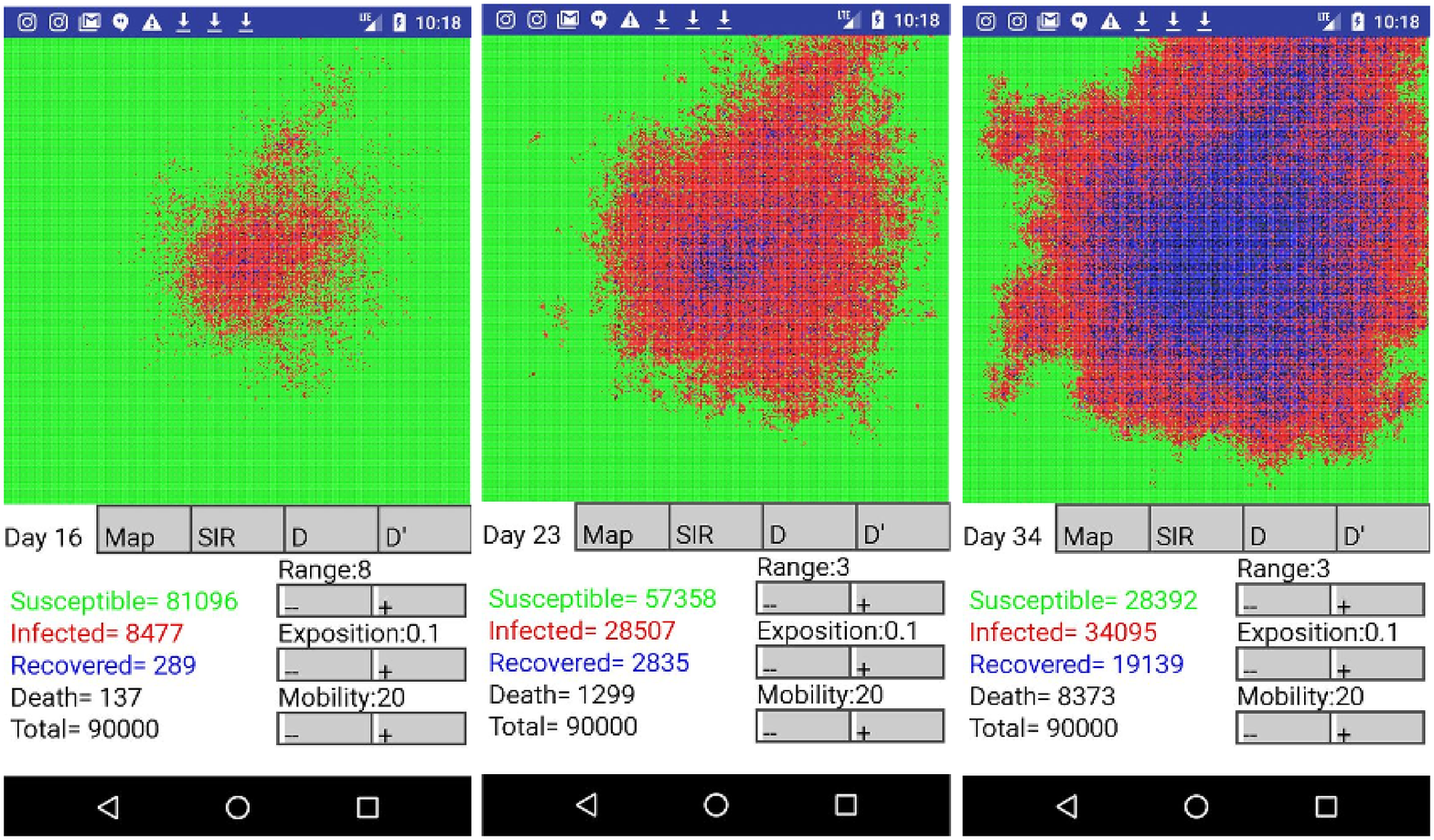}
\caption{Monte Carlo Plank model simulator for Androids illustrating how the pandemic spreads in a lattice of length $L=300$ -- which corresponds to  a  population of $L^2=N=90000$ -- as a function of time.
}
\label{android}
\end{center}
\end{figure*}

Concerning the values of the parameters, the death probability
$p_D=0.3$ is fixed by the total fatalities. In Spain the three
physical parameters are the following. The initial range $R_0=8$,
later changed to $R_0=3$ after the first week to take into account the
lockdown. The temperature or mobility $T=20$, means that in average 20
shots are used to compute the interaction in Eq. (4).  The cross
section multiplied by the density gives the probability of
interaction between cells $\rho\sigma=0.1$. This means about 2
interactions in average. The infection probability is $p_I=0.2$.
Finally the recovery time is $t_r=20.7$ days and the recovery interval
$b=4.5$ days These parameters give an idea of how long the infection
lasts, until there is recovery or death, which is around three weeks.

In Figure 2 we show a simulation of the case of Sweden.  In order to
keep the probability of death equal to the Spanish one, we have
randomly chosen 85\% of the cells with total lockdown, which produces
the same effect as reducing the size of the susceptible ones, in order
to obtain the number of deaths from the first or $ ~ 4500$.  As a
result, the value obtained for the range is larger than in Spain
$R_0=10$, and also the value of $\rho\sigma=0.3$.  In Sweden there
was no lockdown and this value does not change in time.  But in Sweden
there was a gathering restriction that no more than 10 people could
meet. This has been simulated in the Monte Carlo by lowering the
temperature from $T=20$ to $T=15$ during gathering restrictions.  The
gathering day here has been taken equal to 40.
The recovery time $t_r=24$ is about three days larger than in Spain.

In Figure 3 we present results in the case of South Africa.  The first
and second waves are well differentiated and each of them can be
described separately with model D. This is what has been done in the
figure, adjusting the data for each wave separately.  In the same way,
the {\small MCP} model has been adjusted here separately for each of the two
waves. The first Monte Carlo was adjusted for data through August 30,
2020. The second Monte Carlo was adjusted for data from the second
from this date through February 10, 2021.  Each Monte Carlo simulation has
been adjusted independently with a lattice of $L = 300$.  That is why
the probability of death in the first wave is 0.2 whereas in the second
wave is 0.5, since the number of deaths according to these models
would be in a 5:2 ratio between the second and first waves. The
remaining parameters of the Monte Carlo have not changed from those of
the first wave, except for one decimal in the case of the range $R_0$
and the exposition $\rho \sigma$.  The right panel in Figure 3 is a
Monte Carlo prediction of the second wave daily deaths from the data
up to February 10, 2021.

Note the similarity between the parameters of the model for the three
countries that are shown in table 1. This shows some
universality in the physical characteristics that describe the process
in our simplistic model. A more realistic model would require
describing the geographic details and the average population
distribution according to a more realistic geometry than that taken on
a square lattice with constant density. But in view of the results
obtained there is hope that this model will be useful to describe a more
realistic case, but requiring some more work.

To finish, we have also developed~\cite{amaro2} an educative android app to
demonstrate the virus propagation interactively by ploting the daily
grid evolution (Fig. \ref{android}) using a color for each cell state:
susceptible (green), infected (red), recovered (blue) and dead
(black). The resulting {\small SIR} functions S(t), I(t) and R(t) computed with
the {\small MCP} model are also displayed in the app screens, as well as the
daily deaths and cumulative deaths. The initial pandemic parameters
are given by the user in the init screen of the app. 
The Monte Carlo
is started by pressing the running button and the grid evolution is
shown day by day on the screen. Here the user is allowed to change
interactively the parameter range, $R_0$, exposition, $\rho\sigma$, 
 and mobility, $T$, to observe the effect of contention measures and stop the pandemic from spreading. 
 More information, including online lectures and 
the Fortran code, can be found at UWC's nuclear GitHub~\cite{uwcgithub}.

\section{Conclusions}

In this work we have presented a Monte Carlo model ({\small MCP}) of the spread of an
infection, based on interaction parameters inspired by the physics of
many particles. Starting from physical quantities such as interaction
range, cross section and temperature, we translate these concepts
to the
mathematical description of a pandemic under the names of range,
exposition and mobility.

The model consists of individuals in a rectangular lattice that can
interact with their neighbors within a range and at a certain speed,
with a certain probability of interaction, which in turn can produce
infection with the result of recovery or death.

The model fits only for daily deaths, although it could be adapted to
fit other infection data. In this work we have proceeded to calculate
the simplest case with certain simplifications to keep the calculation
as simple and short as possible. In particular, we have set a square
lattice with 90,000 individuals, with which the probability of death
is fixed by the data when they are adjusted to a {\small SIR}-type model.

In general, the model is capable of describing the experimental data
as well as the {\small SIR}-type models, with the difference that the {\small MCP} has a
stochastic behavior, as it is a Monte Carlo simulation subject to
probabilistic parameters. It also allows the inclusion of parameter
modifications over time to simulate mobility restrictions or
quarantines.

The model parameters have been adjusted to describe pandemic waves in
Spain, Sweden and South Africa, with similar results for the
parameters, showing a certain universality of the physical process
involved in the pandemic.

In future work we will study the dependence of the results with the
parameters of the model, as well as extensions to describe larger
populations, of the order of magnitude of complete countries to
determine parameters in more realistic cases.


\section{Acknowledgements}

The authors acknowledge useful discussions with  J\'er\'emie  Dudouet and Ramon Wyss.  
This work is supported by Spanish Ministerio de Economia y
Competitividad and European FEDER funds (grant FIS2017-85053-C2-1-P)
and Junta de Andalucia (grant FQM-225).



\begin{thebibliography}{99}


\bibitem{corona1} D. S. Hui, E. Azhar, T. A. Madani {\it et al.}, The
  continuing 2019-nCoV epidemic threat of novel coronaviruses to
  global health — The latest 2019 novel coronavirus outbreak in Wuhan,
  China. Int. J. Infect. Dis. {\bf 91} (2020) 264-266.


\bibitem{corona2} WHO. Coronavirus disease 2019 ({\small COVID-19})
  Situation Report -- 83. 12 April 2020.
  https://www.who.int/emergencies/diseases/novel-coronavirus-2019/situation-reports.

\bibitem{Gar21} M. Garin {\it et al.}, Epidemic Models for COVID-19 during 
  the First Wave from February to May 2020: a Methodological Review, 
  arXiv:2109.01450.

\bibitem{Coo20a}
I. Cooper, A. Mondal, and  C. G. Antonopoulos, A {\small SIR} model assumption for the spread of COVID-19 in different communities, 
Chaos, Solitons \& Fractals {\bf 139} (2020) 110057.


\bibitem{Coo20b}
I. Cooper, A. Mondal, and  C. G. Antonopoulos,  Dynamic tracking with model-based forecasting for the spread of the COVID-19 pandemic, 
Chaos, Solitons \& Fractals {\bf 139} (2020) 110298.

\bibitem{Lan21} W. Langel,  COVID-19: The Second Wave is not due to Cooling-down in Autumn.
J. Epidemiol. Glob. Health. {\bf 11(2)} (2021) 160.

\bibitem{Pur21}
S. Purkayastha, R. Bhattacharyya, R. Bhaduri {\it et al}., A comparison of five epidemiological models for transmission of SARS-CoV-2 in India, 
BMC. Infect. Dis. {\bf 21} (2021) 533. 


\bibitem{anydist} W. D. Flanders and D. G. Kleinbaum,  Basic Models for
  Disease Occurrence in Epidemiology,  Int. J. Epidemiology {\bf 24} (1995) 1–7.

\bibitem{amaro1} J. E. Amaro, J. Dudouet, and J. N. Orce, Global
  Analysis of the COVID-19 pandemic using simple epidemiological
  models, Applied Mathematical Modeling {\bf 90} (2021) 995–1008.
  
\bibitem{KM} W. O. Kermack and A. G. McKendrick.  A contribution to
  the mathematical theory of epidemics, Proc. Roy. Soc. A {\bf 115} (1927) 700-721.

\bibitem{Wei13} H. Weiss, The {\small SIR} model and the Foundations of Public
  Health, Mater. Mat. no. 3 (2013).

\bibitem{vaccine} S. Chauhan, O. P. Misra, and J. Dhar, Stability analysis of {\small SIR} model with vaccination,  
J. Comput. Appl. Math. {\bf 4(1)} (2014) 17-23.

\bibitem{vaccine2} D. L. Chao and  D. T. Dimitrov, Seasonality and the
  effectiveness of mass vaccination, Math. Biosci. Eng.  {\bf 13(2)} (2016) 249–259.
  
\bibitem{st}  Sashikumaar Ganesan and Deepak Subramani, Spatio-temporal predictive modeling framework for infectious disease spread, 
Sci. Rep. {\bf 11} (2021) 6741.

\bibitem{kendall1957} D. G. Kendall, Discussion of `Measles periodicity and community size' by M. S. Bartlett, J. Roy. Stat. Soc. A {\bf 120} (1957)  64–76.

\bibitem{bartlett1957} M. S. Bartlett, Measles Periodicity and Community Size, J. Royal Stat. Soc. A {\bf 120} (1957) 48-70. 

\bibitem{bartlett1957_2}  M. S. Bartlett, Deterministic and Stochastic Models for Recurrent Epidemics, 
Berkeley Symp. on Math. Statist. and Prob., Proc. Third Berkeley Symp. on Math. Statist. and Prob., Vol. 4, (1956) 81-109.


\bibitem{Tri21}  S. Triambak, D. P. Mahapatra, A random-walk Monte Carlo simulation study of COVID-19-like infection spread.  Physica A {\bf 574} (2021) 126014.

  
\bibitem{nature1} Gang Xie, A novel Monte Carlo simulation procedure for modelling COVID-19 spread over time, 
Scientific Reports {\bf 10} (2020) 13120.  

\bibitem{mc2} S. Maltezos and A.  Georgakopoulou, Novel approach for Monte Carlo simulation of the new COVID-19 spread dynamics.  
Infect. Genet. Evol. {\bf 92} (2021) 104896.



\bibitem{And91} 
R. M. Anderson, Discussion: the Kermack-McKendrick epidemic threshold theorem. Bulletin of mathematical biology, {\bf 53(1)} (1991) 132.

  
\bibitem{Rod16} H. S. Rodrigues, Application of {\small SIR} epidemiological model: new trends, Int. J. Appl. Math. {\bf 10} (2016) 92.   


\bibitem{worldometer}  https://www.worldometers.info/coronavirus/

\bibitem{excessdeaths}  https://www.samrc.ac.za/reports/report-weekly-deaths-south-africa

\bibitem{amaro2}  https://www.ugr.es/\textasciitilde amaro/coronavirus/ 

\bibitem{uwcgithub} https://github.com/UWCNuclear/Covid19\_minischool


\end{thebibliography}
\end{document}